\documentclass[lettersize,journal]{IEEEtran}
\usepackage{amsmath,amsfonts,amssymb}
\usepackage{algorithmic}
\usepackage{algorithm}
\usepackage{array}
\usepackage{eurosym}
\usepackage{bm}
\usepackage{gensymb}
\usepackage{caption}
\usepackage{float}
\usepackage{textcomp}
\usepackage{stfloats}
\usepackage{url}
\usepackage{enumerate}
\usepackage{nccmath} 
\usepackage{verbatim}
\usepackage{graphicx}
\usepackage{cite}
\usepackage{graphicx}
\usepackage{textcomp}
\usepackage{xcolor}
\usepackage{bm}
\usepackage{subcaption}
\makeatletter \renewcommand{\maketag@@@}[1]{\hbox{\m@th\normalsize\normalfont#1}}
\usepackage{subfloat}
\usepackage{balance}
\usepackage{autobreak}

\usepackage{url}
\usepackage{caption}
\usepackage{cite}
\usepackage {epstopdf}
\allowdisplaybreaks[4]
\usepackage{orcidlink} %
\hypersetup{hidelinks} 

\begin{document}
	\title{\huge {Movable Antenna Enhanced Covert Dual-Functional Radar-Communication: Joint Beamforming and Antenna Position Optimization}	\vspace{-3mm}}
	
	\author{Ran Yang$^{\dagger}$, Zheng Dong$^{\star}$, Lin Zhang$^{\dagger}$, Wanting Lyu$^{\dagger}$, \\
		  Yue Xiu$^{\dagger}$, Ning Wei$^{\dagger}$, Ahmad Bazzi$^{\ddagger}$, and Chadi Assi$^{+}$\\
		$^{\dagger}$UESTC, Chengdu, China, $^{\star}$Shandong University, Qingdao, China,
		\\ $^{\ddagger}$NYU Tandon School of Engineering, Brooklyn, USA, 
		 $^{+}$Concordia University, Montreal, Canada,\\
		Emails:  zhengdong@sdu.edu.cn,   wn@uestc.edu.cn,  ahmad.bazzi@nyu.edu, chadi.assi@concordia.ca,\\ \textit{Corresponding Author: Ning Wei\vspace{-12mm}}
		\thanks{This work was supported in part by Mobile Information Networks-National Science and Technology Major Project under Grant 2025ZD1302000, in part by the National Natural Science Foundation of China (NSFC) under Grant 61871070, Grant 61831004, Grant 62203451, and Grant 91938202, in part by Beijing Municipal Science and Technology Program (No. Z24110100120000), in part by China Mobile Communications Group Co.,Ltd, and in part by the Shandong Provincial Natural Science Foundation under Grant ZR2023LZH003.}}
			
		%
		%

	
	
	\maketitle
	\thispagestyle{empty}
	\begin{abstract}
		Movable antenna (MA) has emerged as a promising technology to flexibly reconfigure wireless channels by adjusting antenna placement. In this paper, we study a secured dual-functional radar-communication (DFRC) system enhanced by movable antennas. To ensure communication security, we aim to maximize the achievable sum rate by jointly optimizing the transmit beamforming vectors, receiving filter, and antenna placement, subject to radar signal-to-noise ratio (SNR)  and transmission covertness constraints. To tackle this challenging optimization problem, we first employ a Lagrangian dual transformation process to reformulate it into a more tractable form. Subsequently, the problem is solved by employing a block coordinate descent (BCD) procedure, incorporating semidefinite relaxation (SDR), projected gradient descent (PGD), and successive convex approximation (SCA) techniques. Simulation results demonstrate that the proposed method can significantly improve the covert sum rate, and achieve a satisfactory balance between the communication and radar performance compared with existing benchmark schemes by leveraging the flexibility of movable antennas.
	\end{abstract}
	
	\begin{IEEEkeywords}
		Movable antenna, dual-functional radar-communication, covert communication.
	\end{IEEEkeywords}
	\vspace{-5mm}
	\section{Introduction}
	\IEEEPARstart{D}{ual-functional} {radar-communication  (DFRC) is envisioned to  play a vital role in the upcoming sixth generation (6G) wireless networks~\cite{gonzalez2025six}.  By sharing both spectrum resources and hardware facilities, DFRC has shown great potential in improving system capacity and resource utilization efficiency~\cite{10944644}. However, in DFRC systems, allowing unified probing waveforms to carry private information will pose a high security risk of being wiretapped if the sensing targets are malicious eavesdroppers\cite{9755276}. The inherent broadcast nature of wireless environment further compounds this vulnerability, making the development of efficient security solutions for DFRC systems highly imperative. }

	{Recently, physical layer security (PLS)  and information encryption measures have been explored for DFRC systems~\cite{10608156}. Although these approaches can be employed to protect confidential information from interception, they fail to mitigate the threat to users’ privacy arising from the exposure of transmission behavior~\cite{10090449}. Once the transmission behavior is detected, the adversary may launch subsequent active attacks. For example, it may transmit artificial noise (AN) to disrupt  legitimate links. Although diverse  anti-jamming methods have been developed\cite{11442647}, they generally require prior knowledge of the jammer's transmitter architecture and jamming strategy, which are difficult to obtain in practical hostile environments. In addition, the exposed transmission behavior may facilitate non-computational attacks, such as side-channel analysis, against cryptographic schemes~\cite{7355562}. Thus, to meet the ever-increasing security requirements, covert communication, which shields confidential transmission behaviors from wardens, has been proposed to provide a higher level of security~\cite{7447769}.  Recent research has investigated transmission covertness  in DFRC systems (see \cite{wu2025covert} and references therein). Despite the demonstrated advantages, existing literature  
	mainly focused on conventional fixed-position antennas~(FPAs), while channel variations in the continuous spatial field were not fully exploited. Additionally, the fixed geometric configurations of  FPA arrays can result in array-gain loss during dual-functional beamforming,  thereby constraining overall system performance.}
	
	{To overcome  the limitations of conventional  FPA-based systems, movable antennas (MAs) have recently been proposed as a promising solution to  enhance the dual-task performance~\cite{zhu2025tutorial}. In MA-assisted systems, each antenna element is connected to a radio frequency (RF) chain via flexible cables to support active antenna movement. A prototype of the MA-assisted radar system was initially demonstrated in~\cite{7360379}. Then, channel modeling and performance analysis were explored in~\cite{zhu2023modeling}. Building upon these foundations, a few works have investigated the secure transmission designs for MA-enhanced DFRC systems~\cite{ma2025movable,cao2025joint,le2025beamforming}. However, these works predominantly focus on secrecy rate optimization, and cannot prevent the transmission behavior itself from being detected, thereby limiting their practical effectiveness. }

	To achieve a higher level of transmission security, we investigate the covert transmission design for a movable antenna-enhanced DFRC system. In particular, our contributions are summarized as: 1) We introduce a new secured DFRC design that maximizes the covert sum rate by jointly designing the transmit beamforming vectors, receiving filter, and antenna placement. 2) We develop a block coordinate descent (BCD) procedure for the covert rate maximization problem, incorporating semidefinite relaxation (SDR), projected gradient descent (PGD), and successive convex approximation (SCA) methods. It is worth noting that many subproblems can be optimized in closed forms (or semi-closed forms), which makes the proposed algorithm computationally efficient. 3) We show that MAs can substantially enhance the covert sum rate, and achieve a satisfactory trade-off between the radar performance and communication quality compared to the baseline schemes.
	\section{System Model}
	{We consider a narrowband DFRC system, where  a dual-functional base station (BS) serves $K$ covert users (Bobs) while simultaneously detecting a point-like target.  The BS is equipped with two separate MA-based linear arrays dedicated to signal transmission and reception, respectively. Each array is composed of $N$ MAs. The sensing target is assumed to be a malicious warden (Willie) that attempts  to detect whether the BS is transmitting confidential information. We assume that the feasible movement range for both the transmitting and receiving MAs is a one-dimensional (1D) interval of length $D$. The transceiver antenna positioning vectors (APVs) are denoted by $\boldsymbol{t} = [t_{1},t_{2},\dots,t_{N}]^T\in \mathbb{R}^{N\times 1}$ and $\boldsymbol{r} = [r_{1},r_{2},\dots,r_{N}]^T\in \mathbb{R}^{N\times 1}$, respectively, with $0 \le t_1\le t_2 \dots \le t_{N}\le D$ and $0 \le r_1\le r_2 \dots \le r_{N}\le D$.}
	\vspace{-2mm}
	\subsection{Communication Model}
	Given that the signal propagation  distance is significantly larger than the size of moving regions, the far-field response can be applied for channel modeling\cite{zhu2025tutorial}. Specifically, the angle-of-arrival (AoA), angle-of-departure (AoD), and amplitude of the complex coefficient for each link remain constant despite the movement of  MAs. Note that we adopt the geometric model for communication channels, thus the number of transmission paths at different nodes is the same\cite{zhu2023modeling}.    Denote by $L_k$ the number of transmission paths between the BS and Bob $k$, where the azimuth angle of the $j$-th path at the BS is given by $\psi_{k}^j\in [0,\pi]$.  Then, the signal propagation difference between the position of the $n$-th transmitting MA $t_n$ and the reference point $o^t$ is given by
	$	\rho(t_n,\psi_k^j) = t_n\cos\psi_{k}^j,\forall k,j,n.$ Consequently, the field response vector (FRV) at $t_n$ can be given by 
	\begin{equation}
		\boldsymbol{g}_k(t_n) = \left[e^{\jmath\frac{2\pi}{\lambda}\rho(t_n,\psi_k^1)},\dots,e^{\jmath\frac{2\pi}{\lambda}\rho(t_n,\psi_k^{L_k})}\right]^T\in \mathbb{C}^{L_k\times 1},
	\end{equation} 
	where $\lambda$ is the carrier wavelength. Therefore,  the field response matrix (FRM) of the link from the BS to Bob $k$ for all $N$ transmitting MAs is given by
	\begin{equation}
		\boldsymbol{G}_k({\boldsymbol{t}})\triangleq\left[\boldsymbol{g}_k({t}_1),\boldsymbol{g}_k({t}_2),\dots,\boldsymbol{g}_k({t}_{N})\right]\in \mathbb{C}^{L_k\times N}.
	\end{equation}
	
	{Let $\boldsymbol{\Sigma}_k = \text{diag}\{b_{k,1},b_{k,2},\dots,b_{k,L_k}\}\in\mathbb{C}^{L_k\times L_k}$ denote the path response matrix (PRM), and the channel matrix between the BS and Bob $k$ is given by }
	\begin{equation}
		\boldsymbol{h}_k^H({\boldsymbol{t}}) = \boldsymbol{1}^H\boldsymbol{\Sigma}_k\boldsymbol{G}_k({\boldsymbol{t}})\in \mathbb{C}^{1\times N}, 1\le k \le K.
	\end{equation}
	
	{Denote by $\mathcal{H}_1$ and $\mathcal{H}_0$ the hypotheses that the BS  transmits covert signals or not, respectively. The transmitted signal can be given by}
	\begin{equation} \label{hypotheses}
		\begin{cases} 
			\mathcal {H}_{0}:~\boldsymbol{x}(m) = \boldsymbol{z}_r(m), \;\;\\
			\mathcal {H}_{1}:~\boldsymbol{x}(m) = \sum_{k=1}^{K}\boldsymbol{w}_ks_k(m) + \boldsymbol{z}_r(m).
		\end{cases}
	\end{equation}
	
	{Here, $\boldsymbol{s}(m) = [s_1(m),s_2(m),\dots,s_K(m)]^T\in \mathbb{C}^{K\times 1}$  denotes the communication symbols for $K$ covert users in the $m$-th time slot, $\forall m\in\mathcal{M} = \{1,\dots,M\}$. Meanwhile,  $\boldsymbol{W} = [\boldsymbol{w}_1,\dots,\boldsymbol{w}_K]\in \mathbb{C}^{N\times K}$ denotes the beamforming matrix, and $\boldsymbol{z}_r(m)\in\mathbb{C}^{N\times 1}$ is the dedicated radar signal. It is assumed that $\boldsymbol{s}(m)$ and $\boldsymbol{z}_r(m)$ are independent Gaussian distributed with $\boldsymbol{s}(m)\sim\mathcal{CN}(\boldsymbol{0},\boldsymbol{I}_K)$  and $\boldsymbol{z}_r(m)\sim\mathcal{CN}(\boldsymbol{0},\boldsymbol{R}_s)$, where $\boldsymbol{R}_s \in \mathbb{C}^{N\times N}$ is the covariance matrix of a general rank due to multiple beam transmission. We note that once $\boldsymbol{R}_s$ is determined, the dedicated radar signal $\boldsymbol{z}_r(m)$ can be generated~\cite{4276989}.  Consequently, the covariance matrix of the transmitted signal $\boldsymbol{x}(m)$ can be derived as}
	\begin{equation}
		\begin{cases} 
			\mathcal{H}_0: \boldsymbol{R}_{X}^0 =  \boldsymbol{R}_s,  \;\;\\ \mathcal{H}_1: \boldsymbol{R}_{X}^1 = \sum_{k=1}^{K}\boldsymbol{w}_k\boldsymbol{w}_k^H + \boldsymbol{R}_s. 
		\end{cases}\label{covriance}
	\end{equation} 
	
	{In the proposed system, the quasi-static block fading channels are considered, and the received signal at the $k$-th user under hypothesis $\mathcal{H}_1$ is given by}
	{\begin{align}
			&y_k(m) = \notag\underbrace{\boldsymbol{h}_k^H({\boldsymbol{t}})\boldsymbol{w}_ks_k(m)}_{\text{desired signal}}\\ &  + \underbrace{\sum_{j\neq k}^{K}\boldsymbol{h}_k^H({\boldsymbol{t}})\boldsymbol{w}_js_j(m)+\boldsymbol{h}_k^H({\boldsymbol{t}})\boldsymbol{z}_r(m)}_{\text{mutual interference}} +n_k(m),
	\end{align}}where $n_k(m)\sim \mathcal{CN}(0,\sigma_k^2)$ is the additive white Gaussian noise (AWGN) at the $k$-th user. As such, the signal-to-interference-plus-noise (SINR) at the $k$-th user is given by
	\begin{equation}
		\gamma_k = \frac{|\boldsymbol{h}_k^H({\boldsymbol{t}})\boldsymbol{w}_k|^2}{\sum_{j\neq k}^{K}|\boldsymbol{h}_k^H({\boldsymbol{t}})\boldsymbol{w}_j|^2+\boldsymbol{h}_k^H({\boldsymbol{t}})\boldsymbol{R}_s\boldsymbol{h}_k({\boldsymbol{t}})+\sigma_k^2},
	\end{equation}
	and the achievable rate is given by
	$R_k = \log_2(1+\gamma_k),$ which is also known as the covert rate\cite{10090449}.
	\vspace{-2mm}
	
	\subsection{Radar Model}
	We adopt the line-of-sight (LoS) channel model for the sensing channel between the BS and the target. Let $\varphi$ denote the azimuth angle between  the BS and the target, and the receiving and transmitting steering vectors can be given by 
	$\mathbf{a}_r(\varphi,{\boldsymbol{r}}) = [e^{\jmath\frac{2\pi}{\lambda}\rho({r}_1,\varphi)},\dots,e^{\jmath\frac{2\pi}{\lambda}\rho({r}_{N},\varphi)}]^T$  and $\mathbf{a}_t(\varphi,{\boldsymbol{t}}) = [e^{\jmath\frac{2\pi}{\lambda}\rho({t}_1,\varphi)},\dots,e^{\jmath\frac{2\pi}{\lambda}\rho({t}_{N},\varphi)}]^T$, respectively. Denote by  $\boldsymbol{A}({\boldsymbol{r}},{\boldsymbol{t}}) = 	\mathbf{a}_r(\varphi,{\boldsymbol{r}})\mathbf{a}_t(\varphi,{\boldsymbol{t}})^H$ the response matrix for the sensing target, and the received echo signal is given by
	\begin{equation}
		\boldsymbol{y}(m) = \alpha\boldsymbol{A}({\boldsymbol{r}},{\boldsymbol{t}})\boldsymbol{x}(m)+\boldsymbol{n}_r(m),\label{echo}
	\end{equation}
	where $\alpha$ is the complex reflection coefficient, which captures both the round-trip path loss and radar cross section (RCS) of the target. $\boldsymbol{n}_r(m)\sim\mathcal{CN}(0,\sigma_r^2\boldsymbol{I}_N)$ is the AWGN at the BS. Both communication and dedicated radar waveforms can be exploited as probing signals since they are perfectly known by the BS. Two groups of receiving beamformers should be designed to match the waveforms in~\eqref{hypotheses}. Denote by  $\boldsymbol{u}_i$  the receiving beamformer under $\mathcal{H}_i$, and the corresponding radar SNR can be calculated by
	\begin{equation}
		\text{SNR}_i(\boldsymbol{R}_X^i,\boldsymbol{t},\boldsymbol{r},\boldsymbol{u}_i) =  \frac{|\alpha|^2\boldsymbol{u}_i^H\boldsymbol{A}(\boldsymbol{r},\boldsymbol{t})\boldsymbol{R}_X^i\boldsymbol{A}(\boldsymbol{r},\boldsymbol{t})^H\boldsymbol{u}_i}{\sigma_r^2\boldsymbol{u}_i^H\boldsymbol{u}_i},
	\end{equation} 
	where $i \in \{0,1\}.$ {Note that in this paper, we consider the target tracking stage, in which the sensing target parameters including $\varphi$ and $|\alpha|$ have been estimated in the previous stage and reused for subsequent optimization, rather than being re-estimated in each individual time slot~\cite{9652071}. We assume that the target is quasi-static, so that the parameters remain constant during the entire transmission block and are sufficient for the beamforming design.}
	
	\vspace{-3mm}
	\subsection{Detection Performance and Covertness Constraints}
	In covert DFRC systems, the BS exploits the dedicated radar signal as a cover to achieve covert communication, while the warden Willie seeks to distinguish between the hypotheses in~\eqref{hypotheses} based on the received signals. Let $\boldsymbol{y}_w = [y_w(1),y_w(2),\dots,y_w(M)]^T\in \mathbb{C}^{M\times1}$ be the received signal at Willie, with each element given by
	\begin{equation}
		y_w(m) = \beta\mathbf{a}_t(\varphi,{\boldsymbol{t}})^H\boldsymbol{x}(m)+ n_w(m),\forall m\in\mathcal{M},
	\end{equation}
	where $\beta$ denotes the corresponding path loss, and   $n_w(m)\sim\mathcal{CN}(0,\sigma_w^2)$ is the AWGN at the target. To achieve an optimal test that minimizes detection error probability (DEP) $\xi$,  the likelihood  ratio test is performed  at Willie\cite{10090449}. Specifically, since $\{y_w(m)\},\forall m,$ are independently identical distributed, the probability distribution functions (PDF) of $\boldsymbol{y}_w$ under $\mathcal{H}_1$ and  $\mathcal{H}_0$ can be respectively derived as 
	\begin{align}
		\mathbb{P}_{1} = \mathbb{P}(\boldsymbol{y}_w|\mathcal{H}_1) = \frac{1}{(\pi\eta_{1})^M}\exp\left(\frac{-||\boldsymbol{y}_w||^2}{\eta_{1}}\right),\\
		\mathbb{P}_{0} = \mathbb{P}(\boldsymbol{y}_w|\mathcal{H}_0) = \frac{1}{(\pi\eta_{0})^M}\exp\left(\frac{-||\boldsymbol{y}_w||^2}{\eta_{0}}\right),
	\end{align}
	where $\eta_{0} = |\beta|^2\mathbf{a}_t(\varphi,{\boldsymbol{t}})^H\boldsymbol{R}_s\mathbf{a}_t(\varphi,{\boldsymbol{t}}) + \sigma_w^2$, and $\eta_{1} = \sum_{k=1}^{K}|\beta|^2  |\mathbf{a}_t(\varphi,{\boldsymbol{t}})^H\boldsymbol{w}_k|^2 +\eta_0$, respectively. 
	
	{Denote by $\mathcal{D}_0$ and $\mathcal{D}_1$ the binary decisions in support of $\mathcal{H}_0$ and $\mathcal{H}_1$, respectively. In general, the prior probabilities of hypotheses $\mathcal{H}_0$ and  $\mathcal{H}_1$ are assumed to be equal, and the DEP at Willie can be given by $\xi = P_{MD} +P_{FA},$ where $P_{MD}$ and $P_{FA}$ denote the missed detection probability (MDP) and the false alarm probability (FAP),  respectively. To minimize $\xi$, the optimal likelihood ratio test is adopted \cite{7447769}, which is given by}
	\begin{equation}
		\frac{\mathbb{P}(\boldsymbol{y}_w|\mathcal{H}_1)}{\mathbb{P}(\boldsymbol{y}_w|\mathcal{H}_0)} \overset{\mathcal{D}_1}{\underset{\mathcal{D}_0}{\gtrless}} 1 \Rightarrow ||\boldsymbol{y}_w||^2 \overset{\mathcal{D}_1}{\underset{\mathcal{D}_0}{\gtrless}} \varpi^*,
	\end{equation}
	where $||\boldsymbol{y}_w||^2$ denotes the received signal power, and $\varpi^*$ is the optimal detection threshold, given by 
	\begin{equation}
		\varpi^* \triangleq M\frac{\eta_{0} \eta_{1}}{\eta_{1} - \eta_{0}} \ln \frac{\eta_{1}}{\eta_{0}}.
	\end{equation}
	
	Consequently, the minimum DEP  $\xi^\star$ can be recast as \cite{7447769}
	\begin{subequations} \label{DEP}
		\begin{alignat}{2}
			\xi^\star &= P_{MD} + P_{FA}\\
			&=\mathrm{Pr}(||\boldsymbol{y}_w||^2 \leq \varpi^* | \mathcal{H}_1) + \mathrm{Pr}(||\boldsymbol{y}_w||^2 \geq \varpi^* | \mathcal{H}_0) \\
			&= 1 - \frac{\gamma_{\rm inc}(M,\varpi^*/\eta_0)}{\Gamma_{\rm gam}(M)} + \frac{\gamma_{\rm inc}(M,\varpi^*/\eta_1)}{\Gamma_{\rm gam}(M)},
		\end{alignat}
	\end{subequations}
	{where \( \gamma_{\rm inc} (\cdot, \cdot) \) is the lower incomplete Gamma function given by \( \gamma_{\rm inc} (M, x) = \int_0^x e^{-t} t^{M-1} dt \), and $\Gamma_{\rm gam}(M) = (M-1)!$ is the Gamma function. Thus, the covertness constraint can be given by} 
	\begin{equation}
		\xi^\star \ge 1-\epsilon,\label{dep}
	\end{equation}
    {where $\epsilon\in[0,1]$ denotes the covertness level. Note that a smaller $\epsilon$ corresponds to a more stringent covertness requirement.  However, it is hard to use~\eqref{dep} for further analysis due to the lower incomplete gamma functions. To circumvent this difficulty, we employ  Pinsker's inequality in \cite{7447769} to surrogate $\xi^\star$ by its lower bound, i.e., $\xi^\star \geq 1 - \sqrt{\frac{\mathcal{D}_{\rm KL}(\mathbb{P}_{0}||\mathbb{P}_{1})}{2}},$
    	where $\mathcal{D}_{\rm KL}(\mathbb{P}_{0}||\mathbb{P}_{1})$ is the Kullback Leibler (KL) divergence from $\mathbb{P}_{0}$ to $\mathbb{P}_{1}$, given by}
	\begin{equation}
		\mathcal{D}_{\rm KL}(\mathbb{P}_{0}||\mathbb{P}_{1}) = M\left(\ln\left(\frac{\eta_{1}}{\eta_{0}}\right) + \frac{\eta_{0}}{\eta_{1}} - 1\right).\label{KL}
	\end{equation}
	
	{Combining \eqref{dep} and \eqref{KL}, the covertness constraint is given by }
	\begin{equation}\label{covert1}
		\mathcal{D}_{\rm KL}(\mathbb{P}_{0}||\mathbb{P}_{1}) \le 2\epsilon^2.
	\end{equation}
	
	{Here, we would like to note that since the function $f(x) = \ln x + \frac{1}{x} - 1$ is monotonically increasing for $x\in[1,+\infty)$, the covertness constraint in~\eqref{covert1} can be recast as $	\frac{\eta_{1}}{\eta_{0}} \le \kappa,$ where $\kappa$ is the unique solution of the equation $\mathcal{D}_{\rm KL}(\mathbb{P}_{0}||\mathbb{P}_{1}) =  2\epsilon^2$ in the interval $[1,+\infty)$. }
	\vspace{-2mm}
	\subsection{Problem Formulation}
	In this paper, we aim to maximize the covert sum communication rate by jointly designing the transmit beamforming vectors, receiving filter, and transceiver antenna placement. In particular, the optimization problem can be formulated as  
	\begin{subequations}\label{Problem1}
		\begin{alignat}{2}
			&\underset{ \boldsymbol{W}, \boldsymbol{R}_s,  {\boldsymbol{r}},{\boldsymbol{t},\boldsymbol{u}_0}}{\max} \quad \sum_{k=1}^{K}\log_2(1+\gamma_k) \label{P1_rate}\\
			&\,\text {s.t.}~ \text{SNR}_0(\boldsymbol{R}_s,\boldsymbol{t},\boldsymbol{r},\boldsymbol{u}_0) \ge \Gamma,\label{P1_SNR}\\
			&\hphantom {s.t.~} t_1\ge 0, t_{N}\le D, r_1\ge 0,r_{N}\le D,\label{P1_region}\\
			&\hphantom {s.t.~}  t_n - t_{n-1}\ge d, r_n-r_{n-1}\ge d, 2\le n \le N, \label{P1_distance}\\
			&\hphantom {s.t.~}
			\sum_{k=1}^{K}\boldsymbol{w}_k^H\boldsymbol{w}_k + \text{Tr}(\boldsymbol{R}_s) \le P_t,\label{P1_Power}\\
			&\hphantom {s.t.~}
			\frac{\eta_{1}}{\eta_{0}} \le \kappa,\label{P1_covert}\\
			&\hphantom {s.t.~}
			\boldsymbol{R}_s\succeq 0,\label{P1_R0}
		\end{alignat} 
	\end{subequations} 
	where $\Gamma$ is the radar SNR  threshold\footnote{With $\boldsymbol{R}_s\succeq 0,$ the radar SNR under hypothesis $\mathcal{H}_1$ is guaranteed to be higher than that under hypothesis $\mathcal{H}_0$. Therefore, we focus on sensing SNR performance under $\mathcal{H}_0$ throughout this paper. The optimization on $\boldsymbol{u}_1$ can be performed in a similar fashion as that on $\boldsymbol{u}_0$, and thus omitted  for brevity.}, $d$ represents the minimum distance between MAs to prevent coupling effects, $P_t$ is the total transmission power, and  the constraints in~\eqref{P1_covert} guarantee the covertness level of confidential transmission. We note that the problem in~\eqref{Problem1} is intractable due to the highly non-concave objective function and the coupling of optimization variables.
	\vspace{-4mm} 
	\section{Proposed BCD Algorithm}
	{In this section, we first reformulate the objective function in~\eqref{P1_rate} into a more tractable form by using the Lagrangian dual transformation method. Then, we introduce a BCD-based algorithm, incorporating the SDR, PGD, and SCA methods, the details of which are elaborated as follows.}
	
	\subsection{Problem Reformulation}
	{Based on the Lagrangian dual transformation method~\cite{10772590}, we equivalently transform the original objective function in~\eqref{P1_rate} as}
	\begin{align}
		\mathcal{F}_1(&\boldsymbol{W},\boldsymbol{R}_s,\boldsymbol{t},\boldsymbol{\varrho}) =  \sum_{k=1}^{K}\{\ln (1+\varrho_k)-\varrho_k +\notag\\
		&\frac{(1+\varrho_k)|\boldsymbol{h}_k^H(\boldsymbol{t})\boldsymbol{w}_k|^2}{\sum_{i=1}^{K}|\boldsymbol{h}_k^H(\boldsymbol{t})\boldsymbol{w}_i|^2+\boldsymbol{h}_k^H(\boldsymbol{t})\boldsymbol{R}_s\boldsymbol{h}_k(\boldsymbol{t})+\sigma_k^2}\},
	\end{align}
	{where $\boldsymbol{\varrho} = [\varrho_1,\varrho_2,\dots,\varrho_K]^T\in \mathbb{R}^{K\times1}$ is the slack variable. It can be readily seen that the reformulated objective function is concave with respect to (w.r.t.) $\varrho_k$, and thus the optimal $\varrho_k^\star$ can be derived by checking the first-order optimality  condition, i.e.,}
	\begin{equation}
		\varrho_k^\star = \frac{|\boldsymbol{h}_k^H({\boldsymbol{t}})\boldsymbol{w}_k|^2}{\sum_{j\neq k}^{K}|\boldsymbol{h}_k^H({\boldsymbol{t}})\boldsymbol{w}_j|^2+\boldsymbol{h}_k^H({\boldsymbol{t}})\boldsymbol{R}_s\boldsymbol{h}_k({\boldsymbol{t}})+\sigma_k^2}.\label{varrho_solution}
	\end{equation} 
	
	{Note that with  $\boldsymbol{\varrho}$ being fixed, only the last term of $\mathcal{F}_1(\boldsymbol{W},\boldsymbol{R}_s,\boldsymbol{t},\boldsymbol{\varrho})$, which is in a sum-of-ratio form, is involved in the optimization on $\boldsymbol{W},\boldsymbol{R}_s$ and $\boldsymbol{t}$. To deal with this issue, we first define an auxiliary variable $\boldsymbol{\upsilon} \triangleq [\upsilon_1,\dots,\upsilon_K]^T \in \mathbb{R}^{K\times1}$. Then, the quadratic transformation is employed to recast $\mathcal{F}_1(\boldsymbol{W},\boldsymbol{R}_s,\boldsymbol{t},\boldsymbol{\varrho})$ as $\mathcal{F}_2(\boldsymbol{W},\boldsymbol{R}_s,\boldsymbol{t},\bm{\varrho},\boldsymbol{\upsilon})$, shown at the top of this page.}
	\begin{figure*}[t]
		\begin{equation}
			\mathcal{F}_2(\boldsymbol{W},\boldsymbol{R}_s,\boldsymbol{t},\boldsymbol{\varrho},\boldsymbol{\upsilon}) =  \sum_{k=1}^{K}\{\ln(1+\varrho_k)-\varrho_k+2(1+\varrho_k)\upsilon_k\sqrt{|\boldsymbol{h}_k^H({\boldsymbol{t}})\boldsymbol{w}_k|^2}
			-(1+\varrho_k)|\upsilon_k|^2\left(\boldsymbol{h}_k^H(\boldsymbol{t})\boldsymbol{R}_X^1\boldsymbol{h}_k(\boldsymbol{t})+\sigma_k^2\right)\},\label{F2}
		\end{equation}
			\hrule
	\end{figure*}
 {However, $\mathcal{F}_2(\boldsymbol{W},\boldsymbol{R}_s,\boldsymbol{t},\bm{\varrho},\boldsymbol{\upsilon})$ is still non-concave due to the coupling of optimization variables. Therefore, we  propose a BCD-based  algorithm to obtain an efficient solution.}
	\subsection{Updating Auxiliary Variable}
	With $\boldsymbol{W},\boldsymbol{R}_s,\boldsymbol{t}$ and $\boldsymbol{u}_0$ being fixed, it is observed that  $\mathcal{F}_2(\boldsymbol{W},\boldsymbol{R}_s,\boldsymbol{t},\boldsymbol{\upsilon})$ is concave w.r.t. $\upsilon_k,\forall k$, and the closed-form solution can be given by 
	\begin{equation}
		\upsilon_k^\star = \frac{\sqrt{|\boldsymbol{h}_k^H(\boldsymbol{t})\boldsymbol{w}_k|^2}}{\boldsymbol{h}_k^H(\boldsymbol{t})\boldsymbol{R}_X^1\boldsymbol{h}_k(\boldsymbol{t})+\sigma_k^2},\forall  k.\label{v}
	\end{equation}
	\subsection{Updating Transmit Beamforming}
	{With all other variables being fixed, we focus on optimization on transmit beamforming $\boldsymbol{W}$ and $\boldsymbol{R}_s$. We note that the non-convexity of the problem in~\eqref{Problem1} lies in the quadratic terms w.r.t. $\{\boldsymbol{w}_k\}_{k=1}^K$ in~\eqref{P1_SNR},~\eqref{P1_covert}, and~\eqref{F2}. The SDR method is employed to deal with this issue. Specifically,  we first construct auxiliary variables  $\{\boldsymbol{R}_k\}_{k=1}^K$ with $\boldsymbol{R}_k = \boldsymbol{w}_k\boldsymbol{w}_k^H$, which is a rank-one semidefinite matrix. As such, the covariance matrix $\bm{R}_X^1$ in~\eqref{covriance} can be recast as $\bm{R}_X^1 = \sum_{k=1}^{K}\bm{R}_k + \bm{R}_s.$ Based on~\eqref{F2}, the problem in~\eqref{Problem1} can be recast in the form w.r.t. $\bm{R}_X^1$ and $\{\bm{R}_k\}_{k=1}^K$, i.e., }
	
	\begin{small}
		\begin{subequations}\label{Problem2}
			\begin{alignat}{2}
				&\underset{ \{\boldsymbol{R}_k\}_{k=1}^{K}, {\boldsymbol{R}_X^1}}{\max} ~\sum_{k=1}^{K}\{2(1+\varrho_k)\upsilon_k\sqrt{\boldsymbol{h}_k^H({\boldsymbol{t}}){\boldsymbol{R}}_k\boldsymbol{h}_k({\boldsymbol{t}})}\notag\\
				&\quad\quad\quad-(1+\varrho_k)|\upsilon_k|^2\left(\boldsymbol{h}_k^H(\boldsymbol{t}){\boldsymbol{R}_X^1}\boldsymbol{h}_k(\boldsymbol{t})+\sigma_k^2\right)\} \label{P2_OBJ}\\
				&\,\text {s.t.}~\frac{|\alpha|^2\boldsymbol{u}_0^H\boldsymbol{A}(\boldsymbol{r},\boldsymbol{t})\left({\boldsymbol{R}_X^1}-\sum_{k=1}^{K}\boldsymbol{R}_k\right)\boldsymbol{A}(\boldsymbol{r},\boldsymbol{t})^H\boldsymbol{u}_0}{\sigma_r^2\boldsymbol{u}_0^H\boldsymbol{u}_0}\ge \Gamma,\label{P2_SNR}\\
				&\hphantom {s.t.~}\frac{|\beta|^2\mathbf{a}_t^H(\varphi,\boldsymbol{t}){\boldsymbol{R}_X^1}\mathbf{a}_t(\varphi,\boldsymbol{t}) + \sigma_w^2}{|\beta|^2\mathbf{a}_t^H(\varphi,\boldsymbol{t})\left({\boldsymbol{R}_X^1}-\sum_{k=1}^{K}\boldsymbol{R}_k\right)\mathbf{a}_t(\varphi,\boldsymbol{t})+\sigma_w^2}	\le \kappa,\label{P2_covert}\\
				&\hphantom {s.t.~}{\boldsymbol{R}_X^1} - \sum_{k=1}^{K}\boldsymbol{R}_k\succeq0,~\text{Tr}({\boldsymbol{R}_X^1})\le P_t,\label{P2_power} \\
				&\hphantom {s.t.~} \boldsymbol{R}_k \succeq 0,~\text{rank}(\boldsymbol{R}_k) = 1,1\le k \le K. \label{P2_rank}
			\end{alignat} 
		\end{subequations} 
	\end{small}
	
	{By dropping the rank-one constraints in~\eqref{P2_rank}, the problem in~\eqref{Problem2} is a semidefinite program (SDP), whose solutions $\tilde{\bm{R}}_X^1$ and $\{\tilde{\bm{R}}_k\}_{k=1}^K$  can be directly obtained by using the CVX tool\cite{boyd2004convex}. Here, we would like to note that if  $\{\tilde{\boldsymbol{R}}_k\}_{k=1}^K$ is exactly rank-one, the solution to the relaxed problem is also an optimal solution to the original non-convex problem in~\eqref{Problem2}. While such relaxations are not necessarily tight,  we can always reconstruct a rank-one optimal solution $\{\hat{\bm{R}}_k\}_{k=1}^K$ and the corresponding optimal precoder $\{\hat{\bm{w}}_k\}_{k=1}^K$. Specifically, the rank-one solutions can be reconstructed as follows }
	{	\begin{align}
			&{\hat{\boldsymbol{R}}_X^1} = \tilde{\boldsymbol{R}}_X^1,~{\hat{\boldsymbol{w}}}_k = (\boldsymbol{h}_k^H(\boldsymbol{t})\tilde{\boldsymbol{R}}_k\boldsymbol{h}_k(\boldsymbol{t}))^{-1/2}\tilde{\boldsymbol{R}}_k\boldsymbol{h}_k(\boldsymbol{t}), \\
			&{\hat{\boldsymbol{R}}}_k = {\hat{\boldsymbol{w}}}_k{\hat{\boldsymbol{w}}}_k^H,~ {\hat{\boldsymbol{R}}}_s = {\hat{\boldsymbol{R}}_X^1} - \sum_{k=1}^{K}{\hat{\boldsymbol{R}}}_k.\label{construct}
	\end{align}}

		{Proof: Please refer to Appendix A.}
		\subsection{Updating Transmit Antenna Placement}
		In this subsection, we carry out optimization on $\boldsymbol{t}$. It is worth noting that the non-convexity lies in  the objective function $\mathcal{F}_2(\boldsymbol{t})$ in~\eqref{F2}, and the constraints in~\eqref{P1_SNR} and~\eqref{P1_covert}. To handle this challenge, we introduce a projected gradient descent algorithm, with Nesterov’s acceleration strategy being  incorporated to speed up the convergence\cite{10772590}. Let $\nabla{\mathcal { F}_2}(\boldsymbol{t})\in\mathbb{C}^{N\times 1}$ be the gradient vector at $\boldsymbol{t}$,  and  the antenna position $\boldsymbol{t}$ is updated by the following steps
		\begin{subequations} 
			\begin{alignat}{2}
				&\text{(Step. 1)}~\boldsymbol{m}^{l+1} ={{ {\boldsymbol z}^{l} + \tau^l \nabla  {\mathcal { F}_2}({\boldsymbol z}^{l}) }}, \label{a}\\
				&\text{(Step. 2)}~\boldsymbol{t}^{l+1}  = \text{arg}~ \underset{\boldsymbol{t}}{\min}~ ||\boldsymbol{t}-\boldsymbol{m}^{l+1}||_2^2\notag \\ &~~~~~~~~~~~~~~~~~~~~\text{s.t.}~\eqref{P1_SNR},\eqref{P1_region},\eqref{P1_distance},\eqref{P1_covert}\label{b},\\
				&\text{(Step. 3)}~{\boldsymbol z}^{l+1}  = {\boldsymbol t}^{l+1} + \zeta _{l+1}({\boldsymbol t}^{l+1} - {\boldsymbol t}^{l}),\label{c}
			\end{alignat}
		\end{subequations}
		{where $\boldsymbol{m}^{l+1}\in\mathbb{R}^{N\times1}$ and $\bm{z}^{l+1}\in\mathbb{R}^{N\times1}$ are auxiliary variables at the $(l+1)$-th iteration, and $\tau^l\ge0$ is the descent step length, which can be calculated by the backtracking line search method. Moreover, $\zeta _{l+1} = \frac{q_{l+1}-1}{q_{l+1}}$, and $q_{l+1} = \frac{1+\sqrt{1+4q_l^2}}{2}$ with $q_1 = 0.1.$ }
		
		{For  Step. 1 in~\eqref{a}, we first define 
			that $\tilde{\mathcal{F}}_{i,j}(\boldsymbol{t}) \triangleq \boldsymbol{h}_i^H(\boldsymbol{t})\boldsymbol{R}_j\boldsymbol{h}_i(\boldsymbol{t})$ and $\tilde{\mathcal{W}}_{i}(\bm{t}) \triangleq \boldsymbol{h}_i^H(\boldsymbol{t})\boldsymbol{R}_s\boldsymbol{h}_i(\boldsymbol{t})$, where $1\le i,j\le K$. Thus, the gradient vector $\nabla\mathcal{F}_2(\boldsymbol{t})$ is given by}
		
		\begin{small}
			\begin{align}
				\nabla\mathcal{F}_2&(\boldsymbol{t}) = \sum_{k=1}^{K}(1+\varrho_k)\upsilon_k\frac{1}{\sqrt{\tilde{\mathcal{F}}_{k,k}(\boldsymbol{t})}}\nabla \tilde{\mathcal{F}}_{k,k}(\boldsymbol{t})\notag \\
				&-\sum_{k=1}^{K}\Big((1+\varrho_k)|\upsilon_k|^2\Big)\Big(\sum_{j=1}^{K}\nabla \tilde{\mathcal{F}}_{k,j}(\boldsymbol{t}) + \nabla\tilde{\mathcal{W}}_{k}(\bm{t})\Big),\label{gradient}
			\end{align}
		\end{small}{where $\nabla 	\tilde{\mathcal{F}}_{k,j}(\boldsymbol{t}) \in \mathbb{R}^{N\times 1}$ and $\nabla\tilde{\mathcal{W}}_{k}(\bm{t})\in \mathbb{R}^{N\times 1}$ denote the gradient vectors of $\tilde{\mathcal{F}}_{k,j}(\boldsymbol{t}) $ and $\tilde{\mathcal{W}}_{k}(\boldsymbol{t})$ at $\boldsymbol{t}$, respectively. Please refer to  Appendix B for derivation of corresponding gradient vectors. 
		Then, we move on to deal with the problem in Step. 2. The problem is intractable due to the constraints in~\eqref{P1_SNR} and~\eqref{P1_covert}. To deal with these issues, the SCA method can be employed. According to the second-order Taylor expansion theorem in~\cite{boyd2004convex}, the non-convex parts of constraints in~\eqref{P1_SNR} and~\eqref{P1_covert} can be respectively approximated as}
		
		\begin{small}
			\begin{align}
				\text{SNR}_0(\boldsymbol{t})\geq& \text{SNR}_0(\boldsymbol{t}^l)+\nabla	\text{SNR}_0(\boldsymbol{t}^l)^T({\boldsymbol{t}}-{\boldsymbol{t}}^{l})-\frac{\delta_{0}}{2}||{\boldsymbol{t}}-{\boldsymbol{t}}^{l}||_2^2,\label{P4_SNR_app}\\
				\mathcal{G}(\boldsymbol{t})\leq&\mathcal{G}(\boldsymbol{t}^l) + \nabla \mathcal{G}(\boldsymbol{t}^l)^T (\boldsymbol{t}-\boldsymbol{t}^l)+\frac{\delta_1}{2}||\boldsymbol{t}-{\boldsymbol{t}}^{l}||_2^2\label{P4_covert_app},
			\end{align}
		\end{small}where $\boldsymbol{t}^l$ is the obtained APV in the $l$-th iteration, and  $\mathcal{G}(\boldsymbol{t}) = \mathbf{a}_t^H(\varphi,\boldsymbol{t})\left(|\beta|^2\boldsymbol{R}_X^1-\kappa|\beta|^2\boldsymbol{R}_s\right)\mathbf{a}_t(\varphi,\boldsymbol{t})+(1-\kappa)\sigma_w^2.$ Note that the positive real numbers $\delta_0$ and  $\delta_1$ are selected to satisfy $\delta_0\boldsymbol{I}_N\succeq\nabla^2\text{SNR}_0(\boldsymbol{t})$ and $\delta_1\boldsymbol{I}_N\succeq\nabla^2\mathcal{G}(\boldsymbol{t})$, with $\nabla^2\text{SNR}_0(\boldsymbol{t})$ and $\nabla^2\mathcal{G}(\boldsymbol{t})$ being the Hessian matrices, respectively.  Please refer to the appendix  in \cite{11373884} for the construction of $\delta_0$ and  $\delta_1$. Thus, combining~\eqref{P4_SNR_app} and~\eqref{P4_covert_app}, the problem in~\eqref{b} can be approximated  as follows
		
		\begin{small}
			\begin{subequations}\label{Problem5}
				\begin{alignat}{2}
					&\underset{{\boldsymbol{t}}}{\min} \quad ||\boldsymbol{t}-\boldsymbol{m}^{l+1}||_2^2 \\
					&\,\text {s.t.}~t_1\ge 0, t_N \le D,t_n - t_{n-1}\ge d, 2\le n \le N,\\
					&\hphantom {s.t.~}   \text{SNR}_0(\boldsymbol{t}^l)+\nabla	\text{SNR}_0(\boldsymbol{t}^l)^T({\boldsymbol{t}}-{\boldsymbol{t}}^{l})-\frac{\delta_{0}}{2}||{\boldsymbol{t}}-{\boldsymbol{t}}^{l}||_2^2\ge\Gamma,\\
					&\hphantom {s.t.~}~\mathcal{G}(\boldsymbol{t}^l) + \nabla \mathcal{G}(\boldsymbol{t}^l)^T (\boldsymbol{t}-\boldsymbol{t}^l)+\frac{\delta_1}{2}||\boldsymbol{t}-{\boldsymbol{t}}^{l}||_2^2\le 0,
				\end{alignat} 
			\end{subequations} 
		\end{small}
		which is convex and can be solved by using the CVX tool.
		\vspace{-2mm}
		\subsection{Updating Receive Filter and Antenna Placement}
		{Note that the objective function in~\eqref{F2} is independent of $\boldsymbol{r}$ and $\boldsymbol{u}_0$, which indicates that the receiver design is a feasibility-check problem and the solution will not directly affect $	\mathcal{F}_2(\boldsymbol{W},\boldsymbol{R}_s,\boldsymbol{t},\bm{\varrho},\boldsymbol{\upsilon})$. To provide additional degrees of freedom (DoFs) for optimization on other variables, we propose to maximize the radar SNR for the receiver design, i.e., }
		\begin{subequations}\label{Problem6}
			\begin{alignat}{2}
				&\underset{{\boldsymbol{r}},\boldsymbol{u}_0}{\max} ~ \text{SNR}_0(\boldsymbol{r},\boldsymbol{u}_0) \\
				&\,\text {s.t.}~\eqref{P1_region},\eqref{P1_distance}.
			\end{alignat} 
		\end{subequations} 
		
		As $\boldsymbol{r}$ and $\boldsymbol{u}_0$ are coupled, we likewise employ the BCD algorithm to address this problem. Specifically, with $\boldsymbol{u}_0$ being fixed, the optimization w.r.t. $\boldsymbol{r}$ can be performed similarly  to that of $\boldsymbol{t}$ and is thus omitted for brevity. As for $\boldsymbol{u}_0$, it can be readily observed that the corresponding subproblem is a typical Rayleigh quotient maximization problem. The optimal solution $\boldsymbol{u}_0^\star$ is therefore given by the eigenvector associated with the largest eigenvalue of the matrix $|\alpha|^2 \boldsymbol{A}(\boldsymbol{r}, \boldsymbol{t}) \boldsymbol{R}_s \boldsymbol{A}(\boldsymbol{r}, \boldsymbol{t})^H / \sigma_r^2$.
		\vspace{-3mm}
		\subsection{Convergence and Complexity Analysis}
		{The overall BCD-based algorithm is summarized in \textit{Algorithm 1}. Let $R_{\text{sum}}(\bm{W}^{\iota},\bm{R}_s^{\iota},\bm{t}^{\iota})$ denote the target value of the
			problem in~\eqref{Problem1} at the $\iota$-th iteration, and then we have
			\begin{align} 
				R_{\text{sum}}(\bm{W}^{\iota},\bm{R}_s^{\iota},\bm{t}^{\iota})  \overset {(a)}{=}& \mathcal{F}_1 (\bm{W}^{\iota},\bm{R}_s^{\iota},\bm{t}^{\iota},\bm{\varrho}^{\iota+1}) \notag \\
				\overset {(b)}{=}& \mathcal{F}_2 (\bm{W}^{\iota},\bm{R}_s^{\iota},\bm{t}^{\iota},\bm{\varrho}^{\iota+1},\boldsymbol{\upsilon}^{\iota+1})\notag \\
				\overset {(c)}{\le}&\mathcal{F}_2 (\bm{W}^{\iota+1},\bm{R}_s^{\iota+1},\bm{t}^{\iota},\bm{\varrho}^{\iota+1},\boldsymbol{\upsilon}^{\iota+1})\notag \\
				\overset {(d)}{\le}&\mathcal{F}_2 (\bm{W}^{\iota+1},\bm{R}_s^{\iota+1},\bm{t}^{\iota+1},\bm{\varrho}^{\iota+1},\boldsymbol{\upsilon}^{\iota+1})\notag \\
				\overset {(e)}{\le}&\mathcal{F}_1 (\bm{W}^{\iota+1},\bm{R}_s^{\iota+1},\bm{t}^{\iota+1},\bm{\varrho}^{\iota+1})\notag \\
				\overset {(f)}{\le}&R_{\text{sum}} (\bm{W}^{\iota+1},\bm{R}_s^{\iota+1},\bm{t}^{\iota+1}).
			\end{align} }
			
			{The detailed reasons for these steps are summarized as follows. Steps $(a)$ and $(b)$ hold with equality due to the optimal updates of $\bm{\varrho}^{\iota+1}$ and $\boldsymbol{\upsilon}^{\iota+1}$, respectively. Steps $(c)$ and $(d)$ follow from the monotonic updates of the corresponding subproblems, where the global optimal solutions in~\eqref{Problem2} guarantee $(c)$, and the PGD mechanism ensures $(d)$ by leaving the solution unchanged if no better objective value is found. Finally, steps $(e)$ and $(f)$ arise from the inherent upper-bound properties of the quadratic and Lagrangian dual transformations, respectively. Consequently, the proposed algorithm asymptotically converges to a stationary point\cite{boyd2004convex}.}

		{The computational complexities for optimizing $\boldsymbol{\varrho}$ and $\boldsymbol{\upsilon}$, and updating $\boldsymbol{W}$ and $\boldsymbol{R}_s$ are characterized by $\mathcal{O}(KN^2)$ and $\mathcal{O}(N^{6.5}K^{3.5})$, respectively. Furthermore, the complexities for updating $\boldsymbol{t}$ and $\boldsymbol{r}$ per PGD iteration are $\mathcal{O}(N^{3.5})$, while that for optimizing $\boldsymbol{u}_{0}$ is characterized by $\mathcal{O}(N^3)$. Consequently, the total computational complexity of the proposed algorithm is expressed as $\mathcal{O}\big(I_1(KN^2+N^{6.5}K^{3.5}+(I_2+I_3)N^{3.5}+N^3)\big)$, where $I_1$, $I_2$, and $I_3$ denote the number of iterations required by the outer BCD procedure, and the PGD algorithms for $\boldsymbol{t}$ and $\boldsymbol{r}$, respectively.}
		
		\begin{small}
			\begin{algorithm}[t]
				\caption{BCD Algorithm for the Problem in~\eqref{Problem1}}
				\begin{algorithmic}[1] 
					\STATE \textbf{Initialize}: $\boldsymbol{W}^{\iota},\boldsymbol{R}_s^{\iota},\boldsymbol{t}^{\iota},\boldsymbol{r}^{\iota},\boldsymbol{u}_0^{\iota},\boldsymbol{\varrho}^{\iota},\boldsymbol{\upsilon}^{\iota}$, and set $\iota=0$.
					\REPEAT		
					\STATE Update ${\varrho}_k^{\iota+1}$ via~\eqref{varrho_solution};
					\STATE Update 	$\boldsymbol{\upsilon}^{\iota+1}$ via~\eqref{v};
					\STATE Obtain 	$\tilde{\boldsymbol{R}}_X^{1}$ and $\{\tilde{\boldsymbol{R}}_k\}_{k=1}^K$ by solving the problem in~\eqref{Problem2}, and construct $\{\hat{\boldsymbol{w}}_k^{\iota+1}\}_{k=1}^K$ and $\hat{\boldsymbol{R}}_s^{\iota+1}$ via~\eqref{construct};
					\STATE Update $\boldsymbol{t}^{\iota+1}$  via the PGD algorithm;
					\STATE Update $\boldsymbol{u}_0^{\iota+1}$  via eigenvalue decomposition;
					\STATE Update $\boldsymbol{r}^{\iota+1}$  by the PGD algorithm;
					\STATE Let $\iota = \iota+1$;
					\UNTIL exist conditions are met.
					\RETURN  $\boldsymbol{W}^\star,\boldsymbol{R}_s^\star,\boldsymbol{t}^\star,\boldsymbol{r}^\star$, and $\boldsymbol{u}_0^\star$.
				\end{algorithmic} 
			\end{algorithm}
		\end{small}
		\vspace{-4mm}
		\section{Simulation Results}
		\captionsetup{font={normalsize},labelsep=period}
		\captionsetup[subfloat]{font=normalsize}
		In this section, computer simulations are conducted to  evaluate the performance of the proposed method. We compare our scheme with three baseline schemes: \textbf{1) Upper bound performance scheme}: The optimization is performed to maximize the sum rate without covertness constraints; \textbf{2) Fixed position antenna (FPA)}: The BS is equipped with uniform linear arrays, with $N$ transmitting/receiving antennas spaced between intervals of $\frac{\lambda}{2}$;   \textbf{3) Greedy antenna selection (GAS)}: The moving regions are quantized into discrete ports spaced by $\frac{\lambda}{2}$. The greedy algorithm is employed for the optimization on antenna positions\cite{11373884}.
	\begin{figure}[t]
		\centering
		\includegraphics[width=0.32\textwidth]{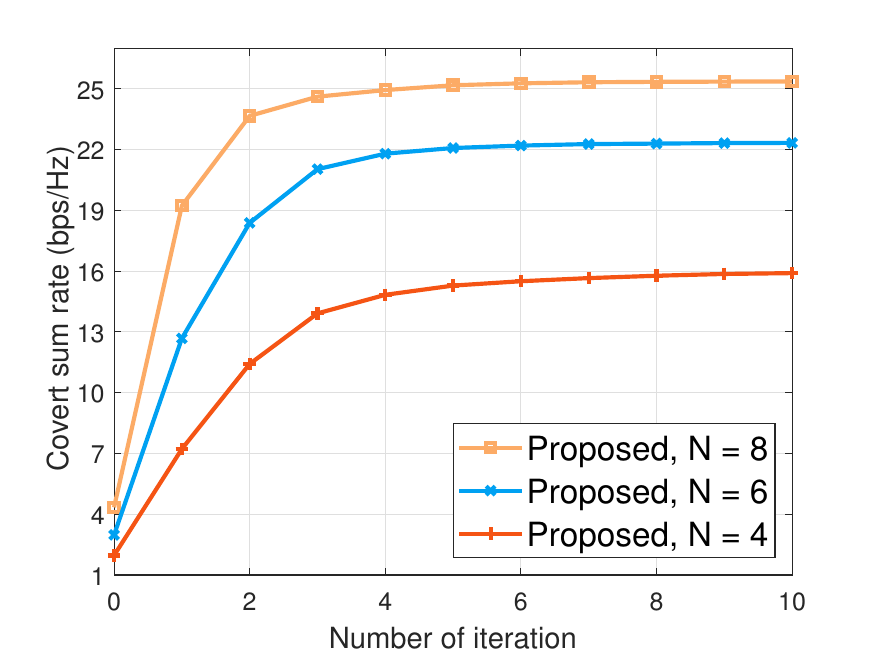}
		\captionsetup{font={normalsize},labelsep=period,singlelinecheck=off}
		\caption{{Convergence behavior.}} 
		\label{iteration} 
		\vspace{-4mm}
	\end{figure}%
		\begin{figure}[t]
		\centering
		\includegraphics[width=0.32\textwidth]{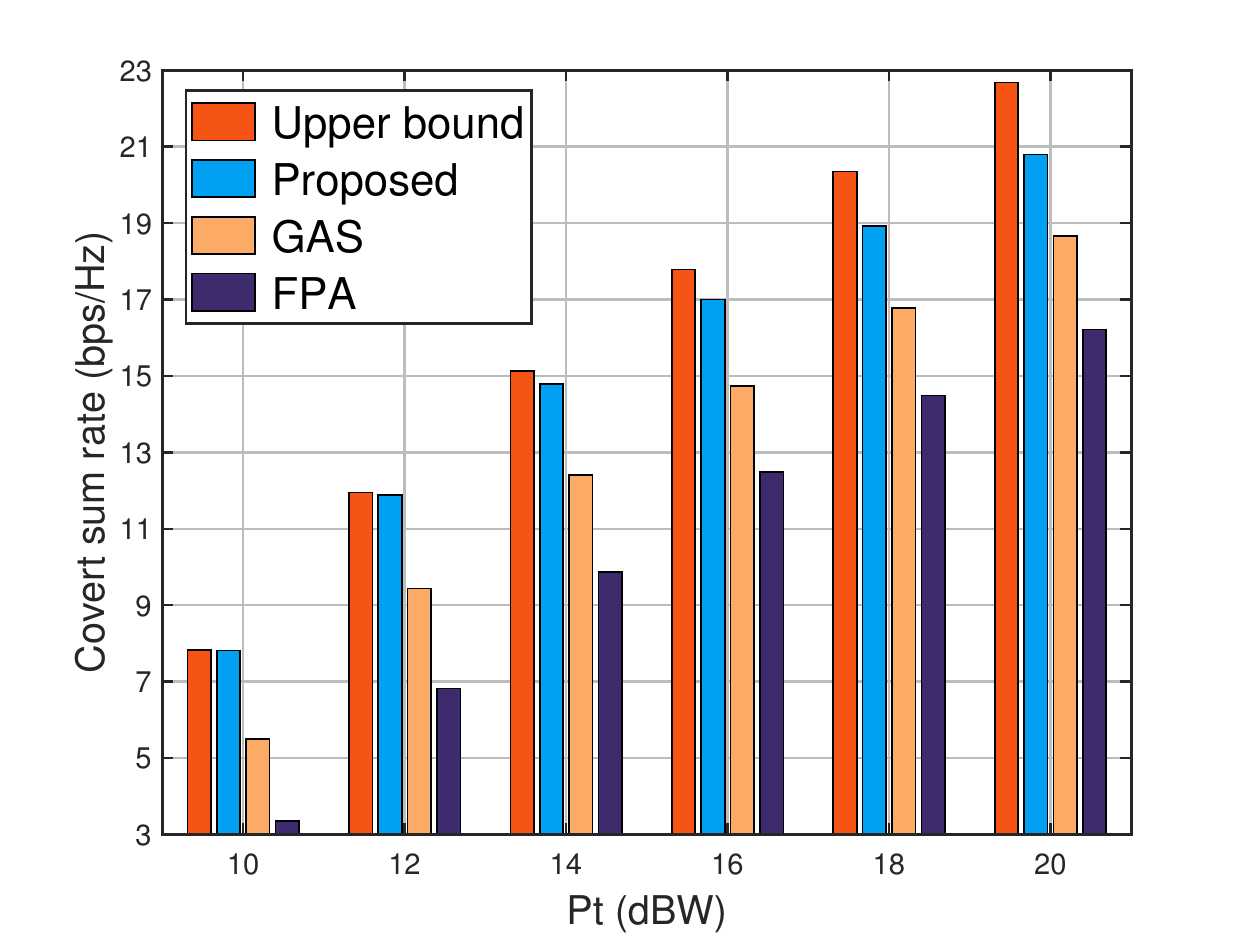}
		\captionsetup{font={normalsize},labelsep=period,singlelinecheck=off}
		\caption{{Covert sum rate versus transmit power $P_t$.}} 
		\label{Pt} 
		\vspace{-4mm}
	\end{figure}%
		\begin{figure}[t]
		\centering
		\includegraphics[width=0.32\textwidth]{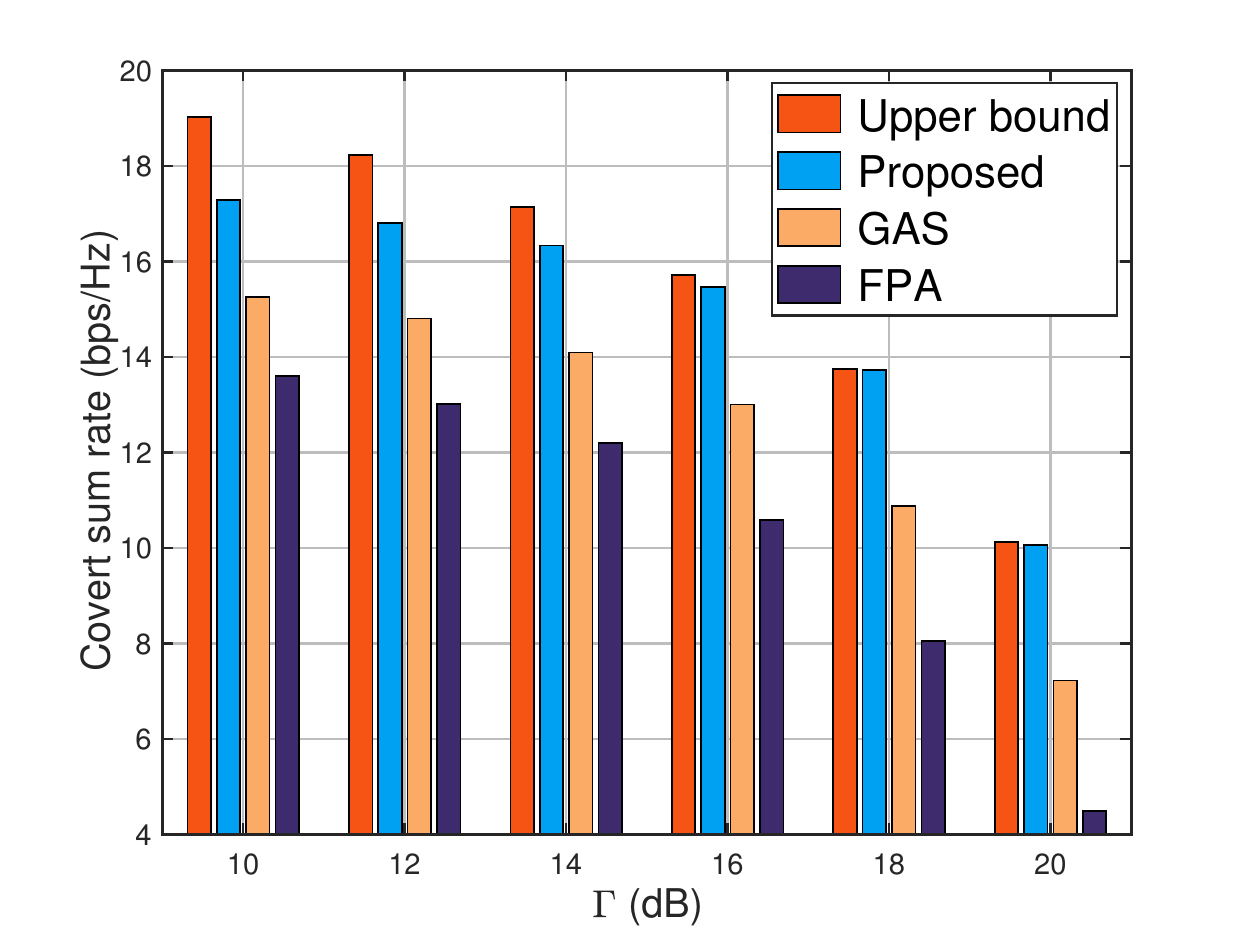}
		\captionsetup{font={normalsize},labelsep=period,singlelinecheck=off}
		\caption{{Trade-off between covert sum rate and radar SNR $\Gamma$.}} 
		\label{SNR} 
		\vspace{-4mm}
	\end{figure}%
	\begin{figure}[t]
		\centering
		\includegraphics[width=0.32\textwidth]{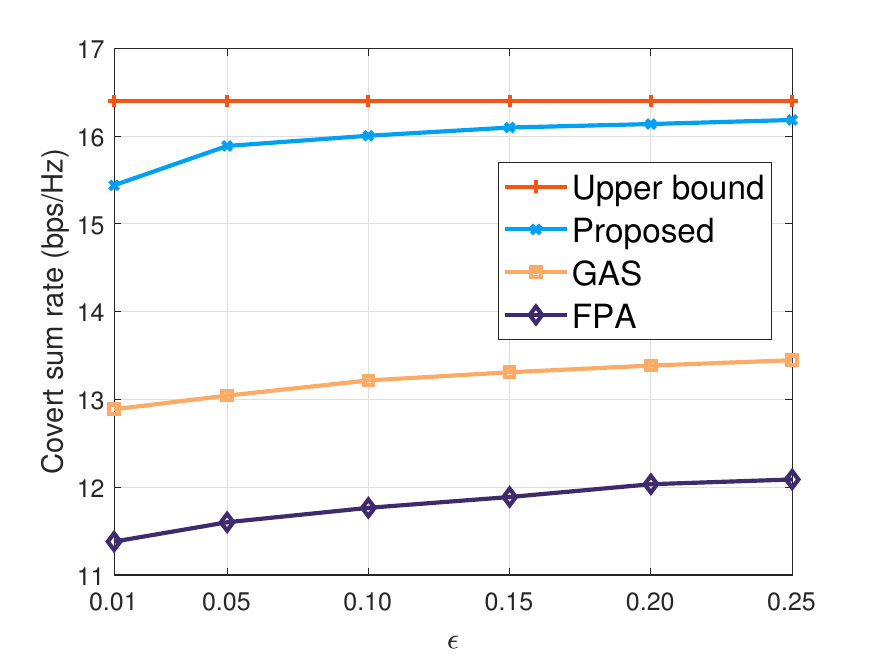}
		\captionsetup{font={normalsize},labelsep=period,singlelinecheck=off}
		\caption{{Covert sum rate versus covertness level $\epsilon$.}} 
		\label{epsilon} 
		\vspace{-4mm}
	\end{figure}%
		\begin{figure}[t]
		\centering
		\includegraphics[width=0.32\textwidth]{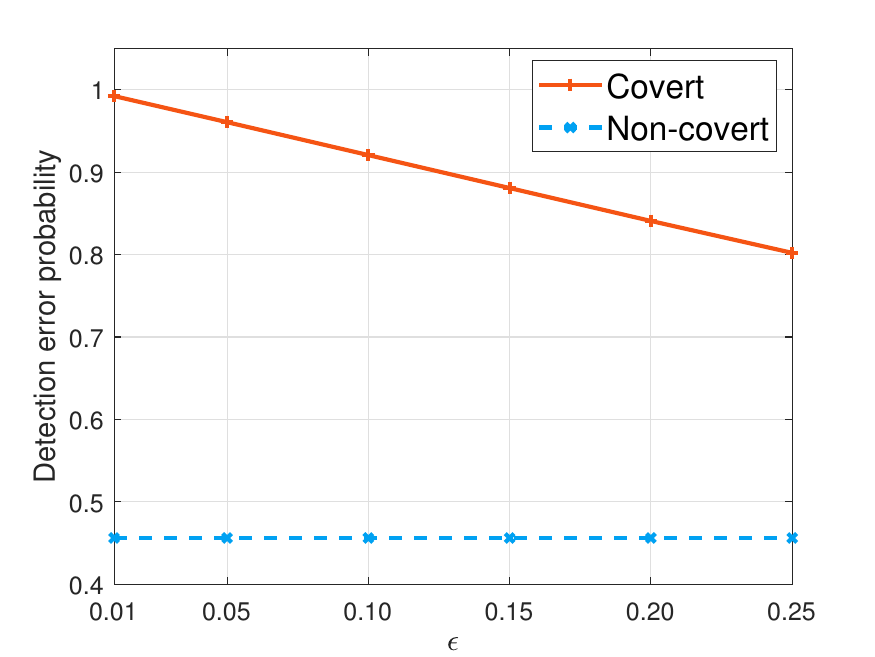}
		\captionsetup{font={normalsize},labelsep=period,singlelinecheck=off}
		\caption{{The DEP versus covertness level $\epsilon$.}} 
		\label{CDF} 
		\vspace{-4mm}
	\end{figure}%
	
		{In our simulation, we assume that the BS is located at (0,~0) m. The users are randomly distributed in a circle centered at (40,~0) m with a radius of 5 m. The numbers of transmitting and receiving paths are identical, i.e., $L_k = L = 6,\forall k$.  The PRM  is given by  $\boldsymbol{\Sigma}_k = \text{diag}\{b_{k,1},\dots,b_{k,L}\}$ with $b_{k,l}\sim\mathcal{CN}\left(0,\frac{c_k^2}{L}\right)$. Note that $c_k^2 = C_0\ell_k^{-\nu}$ denotes the large-scale path loss, where $C_0 = -30~\text{dB}$, and the path-loss exponent $\nu$ is 3.2. Other parameters unless otherwise specified: $K=3,N=4,P_t = 15~\text{dBW}, \Gamma = 15~\text{dB},\varphi = 35\degree, \lambda = 0.1~\text{m},d=\frac{\lambda}{2}$, and $D = 10\lambda.$}
		
		{In Fig.\,\ref{iteration}, we present the convergence behavior of \textit{Algorithm 1}. It can be observed that the covert sum rate monotonically increases and eventually stabilizes as the iterations progress. In most cases, fewer than 8 iterations are sufficient. Note that the number of antennas $N$ has a minor impact on the convergence rate. This is because many subproblems can be optimally solved (e.g., $\boldsymbol{\varrho},\boldsymbol{\upsilon},\boldsymbol{W},\boldsymbol{R}_s$, and $\boldsymbol{u}_0$), which enables the proposed algorithm to quickly converge to a local optimum.}	
		
		{Fig.\,\ref{Pt} illustrates the covert sum rate versus transmission power $P_t$. It can be observed that the proposed scheme significantly outperforms other covert transmission schemes. This superiority can be attributed to the  spatial DoFs provided by MAs, which not only maintain similar PDFs of received signals at the warden under different hypotheses, but also enhance the desired signals, thereby achieving a higher covert rate. Note that the rate gap between the proposed and upper bound performance schemes enlarges with increasing $P_t$, as the latter does not involve transmission covertness, enabling more flexible power allocation.}

		
		{Fig.\,\ref{SNR} demonstrates the communication-sensing trade-off by characterizing the covert rate versus radar SNR threshold~$\Gamma$. In particular, we note that the performance gap between the proposed scheme and the FPA scheme widens with increasing radar SNR $\Gamma$. This phenomenon  can be attributed to the flexibility of antenna movement to achieve a more effective performance trade-off, further highlighting the superiority of MAs.}
		
		{In Fig.\,\ref{epsilon}, we plot the covert sum rate against covertness level $\epsilon$. This result verifies the theoretical analysis that when $\epsilon$ becomes larger, the covertness constraint is looser. Consequently, higher  throughput  can be achieved. We emphasize that the proposed scheme only exhibits a moderate rate degradation compared to the upper bound performance scheme, demonstrating its effectiveness. }
		
		{We further demonstrate the relationship between DEP and covertness level $\epsilon$ in Fig.\,\ref{CDF}. The results show that when $\epsilon$ becomes larger, the DEP of the covert scheme becomes lower. In particular, the DEP of the covert scheme strictly complies with the system's security constraints, whereas the non-covert counterpart invariably fails to do so, thereby validating the effectiveness of the proposed security design.}
		\vspace{-4mm}
		\section{Conclusion}		
		In this paper, we have investigated a movable antenna-enhanced covert DFRC system. A covert sum  rate maximization  problem was formulated by jointly designing beamforming vectors, receiving filter, and transceiver antenna placement. To solve the intractable problem, we developed a BCD-based algorithm, incorporating SDR, PGD, and SCA methods.  Simulation results show that the proposed method can significantly improve the covert sum rate, and achieve a  satisfactory trade-off between the communication and radar performance compared with existing benchmark schemes.
		\vspace{-4mm}
		\appendices
		\section{Proof of Optimality of~\eqref{construct}}
		{First, one can derive that $	
			\boldsymbol{h}_k^H{\hat{\boldsymbol{R}}}_k\boldsymbol{h}_k = \boldsymbol{h}_k^H{\hat{\boldsymbol{w}}}_k{\hat{\boldsymbol{w}}}_k^H\boldsymbol{h}_k = \boldsymbol{h}_k^H\tilde{\boldsymbol{R}}_k\boldsymbol{h}_k,$
			where the APV $\bm{t}$ in $\boldsymbol{h}_k(\bm{t})$ is omitted for notational simplicity. 
			Thus, the value of the objective function  $\mathcal{F}_2(\boldsymbol{W},\boldsymbol{R}_s)$ remains unchanged. Next, we show that $\tilde{\boldsymbol{R}}_k-{\hat{\boldsymbol{R}}}_k\succeq 0.$ For any $\boldsymbol{v}\in\mathbb{C}^{N\times1}$, it holds that $\boldsymbol{v}^H(\tilde{\boldsymbol{R}}_k-{\hat{\boldsymbol{R}}}_k)\boldsymbol{v} = \boldsymbol{v}^H\tilde{\boldsymbol{R}}_k\boldsymbol{v} - (\boldsymbol{h}_k^H\tilde{\boldsymbol{R}}_k\boldsymbol{h}_k)^{-1}|\boldsymbol{v}^H\tilde{\boldsymbol{R}}_k\boldsymbol{h}_k|^2.$
			According to the Cauchy-Schwarz inequality, we have }
		\begin{equation}
			(\boldsymbol{h}_k^H\tilde{\boldsymbol{R}}_k\boldsymbol{h}_k)(\boldsymbol{v}^H\tilde{\boldsymbol{R}}_k\boldsymbol{v}) \ge |\boldsymbol{v}^H\tilde{\boldsymbol{R}}_k\boldsymbol{h}_k|^2.
		\end{equation}
		{So $\boldsymbol{v}^H(\tilde{\boldsymbol{R}}_k-{\hat{\boldsymbol{R}}}_k)\boldsymbol{v}\ge 0$ holds true for any $\boldsymbol{v} \in \mathbb{C}^{N\times1}$, i.e., $\tilde{\boldsymbol{R}}_k-{\hat{\boldsymbol{R}}}_k\succeq 0$. We can conclude that  ${\hat{\boldsymbol{R}}}_s = \hat{\boldsymbol{R}}_X^1-\sum_{k=1}^{K}{\hat{\boldsymbol{R}}}_k\succeq \tilde{\bm{R}}_X^1-\sum_{k=1}^{K}\tilde{\bm{R}}_k = \tilde{\boldsymbol{R}}_s$. Consequently, all the constraints in~\eqref{Problem2} are met. With the derivation above, we can verify that $\{\hat{\boldsymbol{R}}_k\}_{k=1}^{K}$ and $\hat{\bm{R}}_s$ are feasible solutions, and  furthermore, they are globally optimal to the original problem in~\eqref{Problem2}, completing the proof.\hfill$\Box$}
		\vspace{-3mm}
		\section{Derivation of  the Gradient in~\eqref{gradient}}
		Since the derivations of $\nabla 	\tilde{\mathcal{F}}_{i,j}(\boldsymbol{t})$ and $\nabla\tilde{\mathcal{W}}_{i}(\bm{t})$ are similar, for the sake of brevity, we only present the derivation of $\nabla 	\tilde{\mathcal{F}}_{i,j}(\boldsymbol{t})$.
		Recalling that $\boldsymbol{h}_i^H(\boldsymbol{t}) = \boldsymbol{1}^H\boldsymbol{\Sigma}_i\boldsymbol{G}_i(\boldsymbol{t})$,  $\tilde{\mathcal{F}}_{i,j}(\boldsymbol{t})$ can be recast as   $\tilde{\mathcal{F}}_{i,j}(\boldsymbol{t}) = \boldsymbol{a}_i^H\boldsymbol{G}_i(\boldsymbol{t})\boldsymbol{R}_j\boldsymbol{G}_i^H(\boldsymbol{t})\boldsymbol{a}_i$, where $\boldsymbol{a}_i = \boldsymbol{\Sigma}_i^H\boldsymbol{1}\in\mathbb{C}^{L_i\times 1}$. Let us denote the $(n,m)$-th element of $\boldsymbol{R}_j$ as $\boldsymbol{R}_j(n,m) = |\boldsymbol{R}_j(n,m)|e^{\jmath\angle\boldsymbol{R}_j(n,m)}$, and the $l$-th element of $\boldsymbol{a}_i$ as $\boldsymbol{a}_{i}(l) = |\boldsymbol{a}_{i}(l)|e^{\jmath\angle \boldsymbol{a}_{i}(l)}$.
		Thus,  $\tilde{\mathcal{F}}_{i,j}(\boldsymbol{t})$ is recast as 
		{\begin{align}
					\tilde{\mathcal{F}}_{i,j}&(\boldsymbol{t})= \sum_{n=1}^{N}\sum_{l=1}^{L_i}|\boldsymbol{a}_{i}(l)|^2\boldsymbol{R}_j(n,n)
					+\notag\\
					&\sum_{n=1}^{N}\sum_{l=1}^{L_i-1}\sum_{p=l+1}^{L_i}2\mu_{i,j,n,n,l,p}\cos(\phi_{i,j,n,n,l,p})+\notag\\
					&\sum_{n=1}^{N-1}\sum_{m=n+1}^{N}\sum_{l=1}^{L_i}\sum_{p=1}^{L_i}2\mu_{i,j,n,m,l,p}\cos(\phi_{i,j,n,m,l,p}),
				\end{align}	
		}where $\mu_{i,j,n,m,l,p} = |\boldsymbol{R}_j(n,m)||\boldsymbol{a}_{i}(l)||\boldsymbol{a}_{i}(p)|$ and $\phi_{i,j,n,m,l,p} = \angle\boldsymbol{R}_j(n,m)-\angle\boldsymbol{a}_{i}(l)+\frac{2\pi}{\lambda}\rho(t_n,\psi_i^l)+\angle \boldsymbol{a}_{i}(p)-\frac{2\pi}{\lambda}\rho(t_m,\psi_i^p)$. Recalling that $\boldsymbol{t}=[t_1,t_2,\dots,t_N]^T$, the gradient vector $\nabla\tilde{\mathcal{F}}_{i,j}(\boldsymbol{t})$  w.r.t. $\boldsymbol{t}$ is given by $	\nabla\tilde{\mathcal{F}}_{i,j}(\boldsymbol{t}) = \left[ \frac{\partial \tilde{\mathcal{F}}_{i,j}(\boldsymbol{t})}{\partial t_1}, \frac{\partial \tilde{\mathcal{F}}_{i,j}(\boldsymbol{t})}{\partial t_2},...,  \frac{\partial \tilde{\mathcal{F}}_{i,j}(\boldsymbol{t})}{\partial t_N} \right]^T$,
			with each element given by
			
			\begin{small}
				\begin{align}\label{OBJ_app}
					&\frac{\partial\tilde{\mathcal{F}}_{i,j}(\boldsymbol{t})}{\partial t_n} =\notag\\ &\sum_{l=1}^{L_i-1}\sum_{p=l+1}^{L_i}-\frac{4\pi}{\lambda}\mu_{i,j,n,n,l,p}\sin(\phi_{i,j,n,n,l,p})\left(\cos\psi_i^l-\cos\psi_i^p\right)\nonumber\\  
					&+\sum_{m=n+1}^{N}\sum_{l=1}^{L_i}\sum_{p=1}^{L_i}-\frac{4\pi}{\lambda}\mu_{i,j,n,m,l,p}\sin(\phi_{i,j,n,m,l,p})\cos\psi_i^l\notag\\
					&+ \sum_{m=1}^{n-1}\sum_{l=1}^{L_i}\sum_{p=1}^{L_i}\frac{4\pi}{\lambda}\mu_{i,j,m,n,l,p}\sin(\phi_{i,j,m,n,l,p})\cos\psi^p_i.
				\end{align} 
			\end{small}
			
			Thus, the gradient vector $\nabla{\mathcal{F}}_2(\boldsymbol{t})$ can be obtained. \hfill$\Box$
			


			\vspace{-3mm}

			\balance
			\bibliography{reference.bib} 
			\bibliographystyle{IEEEtran} 
			
			%
			
			\vfill
		\end{document}